\RequirePackage{lineno} [2005/11/02 v4.41]
\documentclass[aps,prl,floatfix,footinbib,twocolumn]{revtex4-1}
\usepackage{comment}
\usepackage{epsfig,epstopdf}
\usepackage{amsmath}  
\usepackage[]{nicefrac}
\usepackage{caption}
\usepackage{subcaption}

\newcommand\ket [1] {\ensuremath{\left|#1\right\rangle}}  

\begin{document}

%\setpagewiselinenumbers
%\modulolinenumbers[5]
%\linenumbers

\bibliographystyle{apsrev}

\title{Algorithmic Cooling in Liquid State NMR}

\author{Yosi Atia$^1$, Yuval Elias$^2$, Tal Mor$^3$, Yossi Weinstein$^3$} 
\affiliation{$^1$School of Computer Science and Engineering, The Hebrew University, Jerusalem 91904,  Israel}
\affiliation{$^2$D{\'e}partement IRO, Universit{\'e} de Montr{\'e}al, Montr{\'e}al (Qu{\'e}bec) \  H3C 3J7, Canada}
\affiliation{$^3$Computer Science Department, Technion, Haifa 320008, Israel}

\begin{abstract}

Algorithmic cooling is a method that 
employs thermalization to increase qubit purification 
level, namely it reduces the qubit-system's entropy.
We utilized gradient ascent pulse engineering (GRAPE), 
an optimal control algorithm, to implement algorithmic cooling 
in liquid state nuclear magnetic resonance. 
Various cooling algorithms were applied onto the three qubits
of $^{13}$C$_2$-trichloroethylene, cooling the system beyond 
Shannon's entropy bound in several different ways. In particular, in
one experiment a carbon qubit was cooled by a factor of 4.61. 
This work is a step towards potentially integrating tools of NMR quantum
 computing into in vivo magnetic resonance spectroscopy.
\end{abstract}

\maketitle
\def \eps {\varepsilon}

\section{Introduction}
The quantum computational model permits algorithms that provide significant --- and sometimes even exponential ---
speed-up over any known classical 
counterpart~\cite{Simon94,Shor97,HHL09}. 
A rather different scope of that model is to enable improved quantum
technologies, e.g. quantum repeaters for communicating secure data over longer distances~\cite{BDCZ98}.
Algorithmic cooling, experimentally implemented in this work, is a method that 
might contribute to both scopes. On the one hand, it was originally 
suggested as a method for increasing the qubits' purification 
level~\cite{BMRVV02,FLMR04,SMW05,RM15,R-BL15,PR-B+15}, as qubits in a highly pure state are required
both for initialization and for fault tolerant~\cite{KLZ98, AB97} quantum 
computing. On the other hand, the suggested novel usage of data 
compression may potentially be found useful for
increasing the signal to noise ratio of liquid-state NMR and
in vivo magnetic resonance 
spectroscopy~\cite{FLMR04,AAC-pat,EGMW11}.

Nuclear magnetic resonance quantum computing 
(NMR-QC)~\cite{CFH96,CLK+00,Glaser01,VC05,Jones11} commonly uses spin \nicefrac{1}{2} nuclei 
(hereinafter \emph{spins}) of molecules as qubits. 
Compared to other implementations of small quantum computing devices, 
liquid-state NMR has an advantage of relatively easy realization of 
quantum gates by applying RF fields and utilizing spin-spin interactions. 
However, NMR-QC also has some disadvantages due to working with an ensemble of
spins in a mixed state~\cite{GC97, BMRV10}, e.g. it is not scalable. Algorithmic
cooling, in theory, resolves that problem~\cite{BMRVV02,FLMR04,SMW05}.

The thermal energy at room temperature is much higher than the 
magnetic potential energy of nuclear spins even in the most advanced NMR 
devices. 
Therefore, at equilibrium, the qubit ensemble is in a highly mixed state 
- the probability difference between the ``up'' and ``down'' states 
(hereinafter the \emph{polarization}, denoted as $\eps$) is very small. At thermal equilibrium
\begin{equation}
\eps=P_{\uparrow}-P_{\downarrow}=\tanh\left(\frac{\Delta E}{2k_{\rm B}T}\right)\xrightarrow[\Delta E \ll k_{\rm B}T]{}\frac{\Delta E}{2k_{\rm B}T}=\frac{\hbar\gamma B_{z}}{2k_{\rm B}T}.
\label{eq:equlibrium_pol}
\end{equation}
\\
Here $\gamma$ is the gyromagnetic ratio of the spin, 
$B_z$ is the intensity of the magnetic field, 
$k_{\rm B}$ is the Boltzmann constant and $T$ is the bath temperature. 
When outside of equilibrium, spins with higher polarization than their thermal equilibrium 
polarization are considered ``cool'', and we can use Eq.~\ref{eq:equlibrium_pol} to define the spin temperature as $T_{\rm spin}=\frac{\hbar\gamma B_z}{2k_{\rm B}\eps}$.

Upper bounds on spin cooling (i.e. on polarization enhancement) 
can be derived by interpreting the spin state in terms of information 
theory~\cite{CT06}. 
The information content ($IC$) of the spin was defined using the conventional
 notion of Shannon entropy $H$. 
The relation between a single spin's polarization and $IC$ 
is given by the following equation~\cite{POTENT,EFMW07}
\begin{equation}
\begin{split}
H_{1qubit}=&\biggr[\frac{1-\eps}{2}\ln\left(\frac{1-\eps}{2}\right) +\frac{1+\eps}  {2}\ln\left(\frac{1+\eps}{2}\right)\biggr]\\
IC_{1qubit}=&1-H_{1qubit}=\frac{\eps^{2}}{\ln4}+O(\eps^4).
\end{split}
\end{equation}
The information content of a spin system is invariant to reversible operations,  and therefore bounds the maximal $IC$ a single spin can reach by lossless 
manipulations, such as quantum gates. 
This entropy bound, also often called \emph{Shannon's bound}, 
limits the maximal polarization of a single spin, 
given an initial thermal state of the spin system.

\begin{figure} 
	\centering
	\includegraphics [scale=0.45] {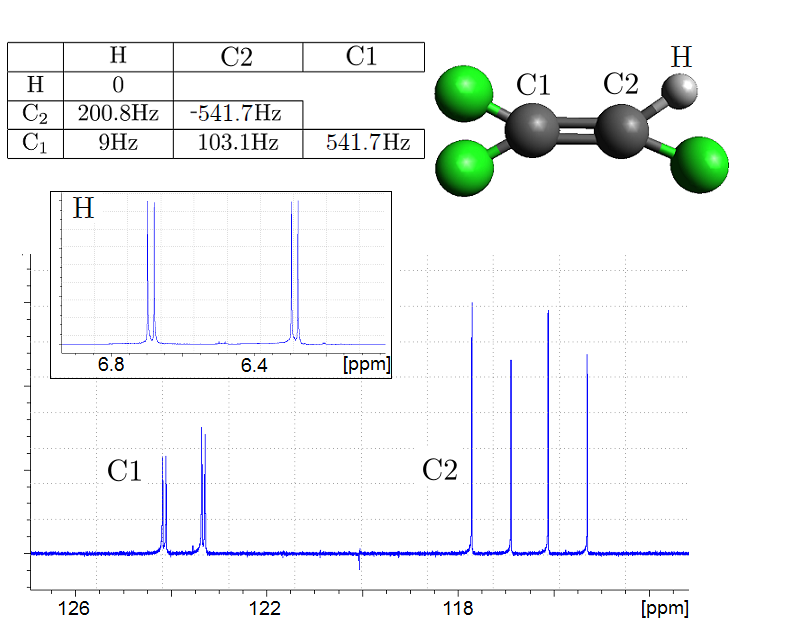}		
	\caption{(Color online) $^{13}$C$_2$-TCE  with paramagnetic reagent Cr(acac)$_3$, in CDCl$_3$ (chloroform-d) solution. The experiments were performed on a Bruker Avance III 600 spectrometer using a standard 5~mm double resonance probe with a
broadband inner coil tuned to 13C and an outer 1H coil probe. This sample has three active spins marked H, C2, and C1. In the table, the chemical shifts relative to the transmitter frequency are in the diagonal cells, and the J-couplings are in the off-diagonal cells. The carbon spectrum is at the bottom, the proton spectrum is in the small frame. The units of the x axis are parts-per-million.} 
\label{fig:TCE_with_spectrum}
\end{figure}

In our spin system, $^{13}$C$_2$-trichloroethylene (TCE,  see Figure~\ref{fig:TCE_with_spectrum}), consisting of a proton and two $^{13}$C, the $IC$ at thermal equilibrium is:
\begin{equation}
IC_{eq}=\frac{\eps_{H,eq}^2}{\ln4}+2\frac{\eps_{C,eq}^2}{\ln4}=\left[\frac{\gamma^2_H}{\gamma^2_C}
+ 2\right]\cdot\frac{\eps_{C,eq}^2}{\ln4}=17.84\frac{\eps_{C,eq}^2}{\ln4} \ .
\end{equation}
Shannon's bound dictates that a single spin cannot hold more than the equilibrium information content of the entire spin system:
\begin{equation}
\begin{split}
IC_{eq}=17.84&\frac{\eps_{C,eq}^2}{\ln4}=\frac{\eps_{max}^2}{\ln4}\\
\Rightarrow \eps_{max}&=4.224\eps_{C,eq}.
\end{split}
\label{eq:maxcooling}
\end{equation}
For convenience, we approximate 
$\gamma_H/\gamma_C =4$, and then $\widetilde{IC}_{eq}=18$,
and 
$\widetilde{\eps}_{max}=4.24\eps_{C,eq}$.

Algorithmic Cooling (AC) of spins counter-intuitively utilizes the heat bath, 
that decays polarizations to thermal equilibrium, to cool spins beyond Shannon's bound. 
AC requires  a spin system where some spins, called \emph{reset spins}, 
thermalize significantly faster than other spins, called \emph{computation spins}. 
Reversible polarization compression (hereinafter \emph{compression}) is applied 
to the spin system, transferring some of the computation spins' entropy 
to the reset spins, which quickly lose some of it to the environment.
The process can be repeated, converging the system to a stable trajectory (limit-cycle) 
in the thermodynamic diagram. 
The efficiency and the cooling limit of AC are (ideally) dependent 
on the unitary restriction of processes between reset steps, and on
the ratio between the relaxation times 
of the cooled spins and the reset spins.
 
Various cooling algorithms were developed, following the basic principle
presented in~\cite{BMRVV02}. 
For example, in a three qubit system with uniform equilibrium polarization $\eps$, 
the initial information content is $IC_{eq}=3{\eps^2}/{\ln4}$,   
the maximum polarization of a single spin that can be reached using unitary 
transformations~\cite{Sorensen89,SV99},
is $1.5\eps$,  
and Shannon's bound for the maximal polarization of a single spin is
$\sqrt{3}\eps = 1.732 \eps$. 
But if one spin has a much shorter thermalization time than the other spins, 
it will reset while the others retain most of their polarization, 
so that the entire spin system is cooled. 
Ideally, iterating the compression process twice leads 
to a bias of $1.75\eps$, bypassing the result obtained by unitary transformations 
and even bypassing 
Shannon's bound~\cite{FLMR04}. 
Repeating the process while assuming infinite relaxation 
time ratios allows enhancing the polarization of one spin asymptotically
to 2$\eps$~ \cite{Jose-PhD-Thesis}. 
Algorithms applying these processes to~$n$ qubits ideally cool exponentially 
beyond the unitary cooling~\cite{BMRVV02,FLMR04}, and can be  
practicable or optimal, see~\cite{FLMR04,SMW05,EFMW06,EFMW07,SMW07,EMW11}.

In TCE, the proton reset spin has higher equilibrium polarization 
than C1 and C2, the $^{13}$C computation spins. 
In such scenarios, even a special case --- AC without compression 
(called heat bath cooling~\cite{POTENT,EGMW11}), can cool 
the spin system beyond Shannon's bound. This can be 
done by applying a polarization transfer~\cite{MF79} from the proton to C1, or alternatively, by swapping the two polarizations via a polarization
 exchange (PE) gate,
and waiting for the proton to regain some of its 
polarization (while the carbon is still cool). 
A successive PE from the proton to C2 followed by another waiting period yields polarization of approximately $4$  
on all three spins,
in units of carbon equilibrium polarization.  
If the relaxation time ratio is sufficiently large, and all gates are perfect 
then $\widetilde{IC}_{total} \rightarrow 48$, far above the initial approximate value of $18$. Further cooling can be achieved using compression. 

In practice, heat bath cooling  
of TCE~\cite{POTENT}, 
yielded polarizations $\{1.74, 1.86, 3.77\}$ 
for C1, C2 and the proton respectively, 
well below the ideal
$\{4,4,4\}$. 
Yet, the resulting total 
$IC$ is $20.70$ ($\pm 0.06$), which is beyond
(and statistically significant) the experimental initial $IC$ 
(of $17.84$) at equilibrium,
hence showing for the first time that the Shannon bound can be experimentally
bypassed. 
Heat bath cooling on 
two amino-acids~\cite{EGMW11} also successfully
bypassed Shannon's bound later on.
On the other front, experimental work, cooling solely by compression was done 
by S{\o}rensen~\cite{Sorensen89} on methylene chloride, and by 
Chang, Vandersypen and Steffen~\cite{CVS01}  
on three fluorines in C$_2$F$_3$Br. 
Full AC~\cite{BMR+05} and multi-cycle AC~\cite{RMBL08} 
using solid-state NMR was successfully done at the University of Waterloo.

\section{Materials and methods}
\begin{figure} 
	\centering
	\includegraphics [width=\columnwidth] {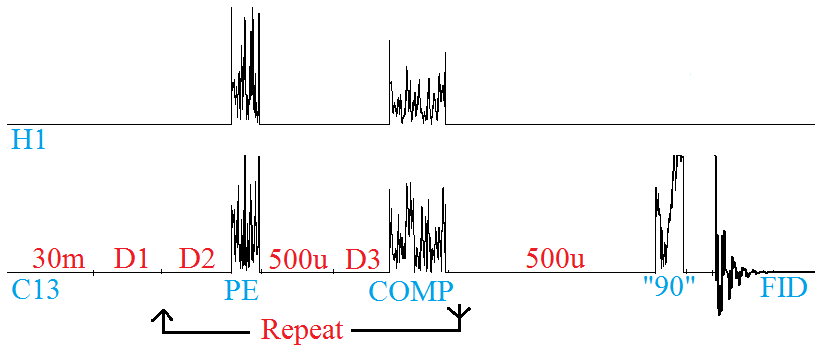}		
	\caption {(Color online) Visualization of Process 1, based on the TopSpin$^\copyright$ output, 
showing the RF power vs time in both channels, followed by an aquisition of the free induction decay (FID). 
The PE and COMP pulses are 6.5~msec and 13~msec long respectively, 
and the delays maximizing the polarization of C1 are D2=5s, D3=3s. 
The delay D1 was set to 150s to equilibrate the system.} 
			
\label{fig:BrukerACscheme}
\end{figure}

\begin{figure} 
	\centering
	\includegraphics [width=\columnwidth] {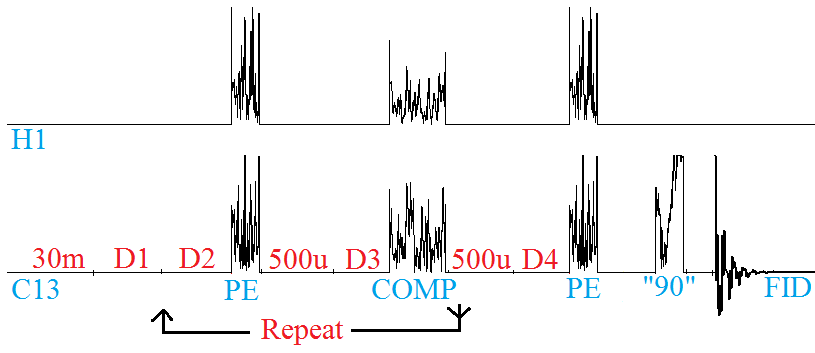}		
	\caption {(Color online) Visualization of Process 2, based on the TopSpin$^\copyright$ output, 
showing the RF power vs time in both channels, followed by an aquisition of the free induction decay (FID). 
 The delays maximizing the polarization of C1 are D2=5s, D3=3s, and D4=5s.} 
			
\label{fig:Process2}
\end{figure}

\begin{figure} 
	\centering
	\includegraphics [width=\columnwidth] {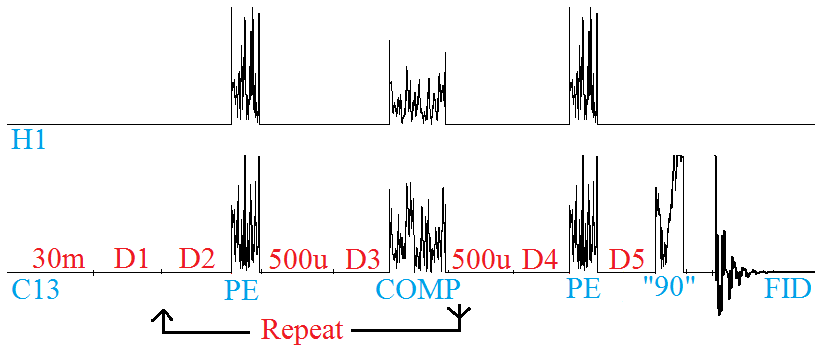}		
	\caption {(Color online) Visualization of Process 3, based on the TopSpin$^\copyright$ output, 
showing the RF power vs time in both channels, followed by an aquisition of the free induction decay (FID). 
The delays maximizing the polarization of C1 are D2=5s, D3=3s, D4=6s, and D5=6s. } 
			
\label{fig:Process3}
\end{figure}

In order to implement AC and multiple-cycle AC on liquid-state TCE 
we utilized (following~\cite{RMBL08})
Gradient Ascent Pulse Engineering  
(GRAPE)~\cite{KRK+05}, an optimal control algorithm, 
to generate high fidelity pulses
for obtaining the compression gate 
and the PE gate~\cite{AEMW14}.
Here we present various algorithms for cooling liquid TCE. Process 1 (see Figure \ref{fig:BrukerACscheme}), aimed to maximize $IC_{C1}$, is as follows:
\begin{enumerate}
\item Wait for duration D2 (H regains polarization)
\item PE(H$\rightarrow$ C2)
\item Wait for duration D3 (H regains polarization)
\item Compression of C1,C2,H onto C1.
\item Return to step 1, unless C1 is saturated.
\end{enumerate}
Ideally, the polarization of C1 saturated at $\widetilde{IC}_{C1}\rightarrow 64$.  Process 2, aimed to maximize $IC_{C1,C2}$ is composed of Process 1 followed by a wait step for duration D4, and by PE(H$\rightarrow$ C2) to cool C2, ideally reaching $\widetilde{IC}_{C1,C2}=80$ (see Figure~\ref{fig:Process2}). The goal of Process 3 is to maximize $IC_{total}$, hence we apply Process 2, followed by a wait step for duration D5, ideally reaching $\widetilde{IC}_{total}=96$ (see Figure~\ref{fig:Process3}).
In all cases, a read-out pulse was applied on the spin of interest prior to acquisition.

In the experiment, the measured relaxation times 
(see table ~\ref{tbl:TCET1T2}),  
were obtained by inversion recovery as in~\cite{POTENT,AEMW14}.
Adding 
a paramagnetic reagent to the TCE, improved the relaxation 
time ratios as suggested in~\cite{FMW05}. 
We simulated the three processes using the experimental delays and
measured relaxation times, while assuming perfect pulses. According to the simulation,
the polarization of C1 could be enhanced 
by a factor of 5.49 after seven rounds ($IC_{C1}=30.13$, see also Figure \ref{fig:buildup})
via Process 1, the polarization of the two carbon spins 
could reach 4.78 and 3.70 ($IC_{C1,C2}=36.53$) via Process 2.
The polarization of three spins could reach $\{3.98, 2.97, 3.75\}$ ($IC=38.73$) via Process 3.


The implemented PE and compression pulses were generated using SIMPSON 
version 3.0~\cite{SIMPSON+00,SIMPSON+08}, an open source program 
implementing GRAPE. The pulses were designed to be robust 
to deviations up to $\pm$15\% in RF power~\cite{AEMW14}. The pulses were not designed to apply a specific unitary gate, but to apply a less constraining  state-to-state transformation. However the state of the system changes with each cooling cycle. Therefore, among two pulses that apply PE, even though one pulse performs better in equilibrium~\cite{AEMW14}, we used another pulse, which yielded better cooling for the entire process.  
Let $\{\{\dots\}\}$ stand for the diagonal of a density matrix in the
computation basis. 
The PE pulse was designed to evolve the system from equilibrium 
\begin{equation}
 4I_z^H+I^{C2}_z+I^{C1}_z \propto \{\{6,4,4,2,-2,-4,-4,-6\}\} \ ,
\end{equation}
to a finite state, 
\begin{equation}
 I_z^H+4I^{C2}_z+I^{C1}_z \propto \{\{6,4,-2,-4,4,2,-4,-6\}\} \ ,
\end{equation}
where $I_z = \frac{1}{2} \sigma_z$. 
The two states are represented 
here~\footnote{In the actual design we used the more precise $\gamma$ ratio
of 3.98, but in the explanation here we use 4 for clarity.} 
in product operator formalism as the reduced 
(shifted and scaled~\cite{EFMW06}) diagonal density operator~\cite{CFH96,AEMW14}.
The compression pulse (COMP) was designed to evolve the system from 
\begin{equation}
 I_z^H+I^{C2}_z+I^{C1}_z \propto \{\{3,1,1,-1,1,-1,-1,-3\}\}
\end{equation}
(a state of three spins with identical polarizations)
to 
\begin{equation}
\begin{split}
 \frac{1}{2}I_z^H+&\frac{1}{2}I^{C2}_z+\frac{3}{2}I^{C1}_z+2I_z^HI^{C2}_zI^{C1}_z \\
 \propto & \{\{3,1,1,1,-1,-1,-1,-3\}\}.
\end{split}
\end{equation}
We chose this final state, as the four highest probabilities correspond 
to the four states where C1's spin is 0, namely, $\ket{0ij}, ij\in{0,1}$. 
Notice that the polarization-increase factor of C1 is 1.5, the maximum 
possible under unitary transformations, as mentioned above.

\section{Results}
\begin{figure} 
	\centering
	\includegraphics [scale=0.36] {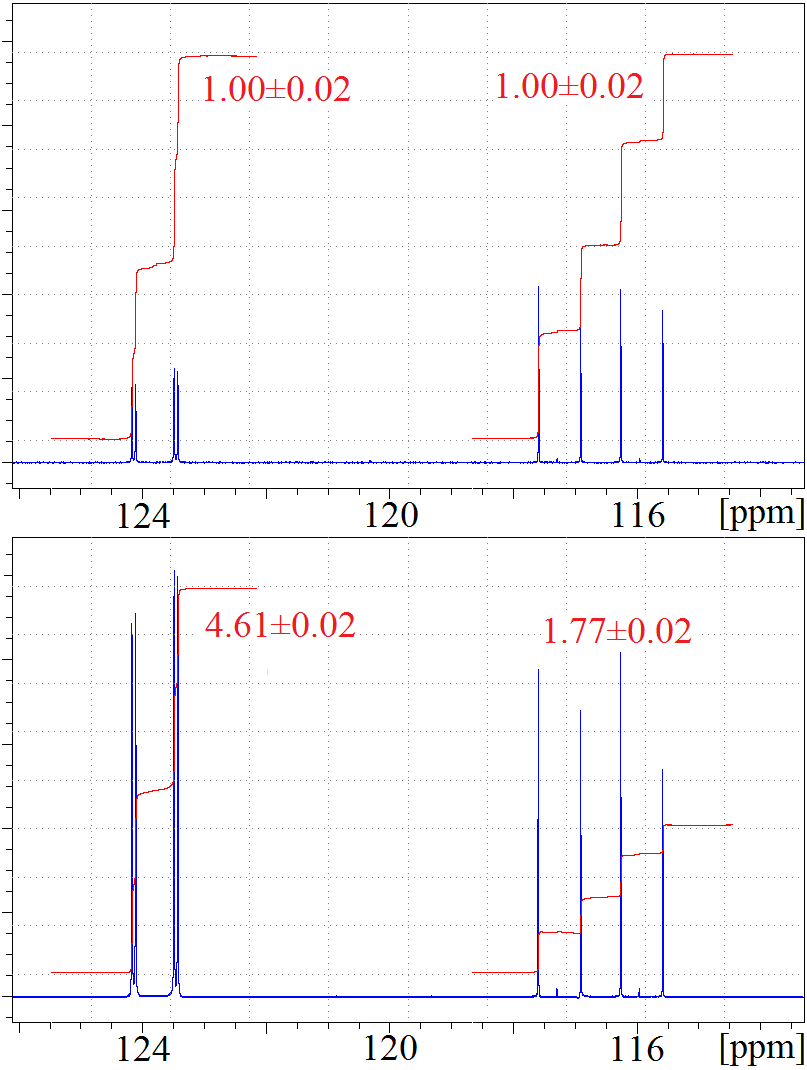}		
	\caption {(Color) the $^{13}$C carbon spectrum before and after seven cycles of algorithmic cooling to maximize the polarization of C1. The peak integrals are displayed in red (arbitrary units).}
			
	\label{fig:Cspectrumbefore+after}
\end{figure}

\begin{figure*}
\centering
\begin{subfigure} [b] {\columnwidth}
C1 Information Content Buildup
\includegraphics [width=\columnwidth] {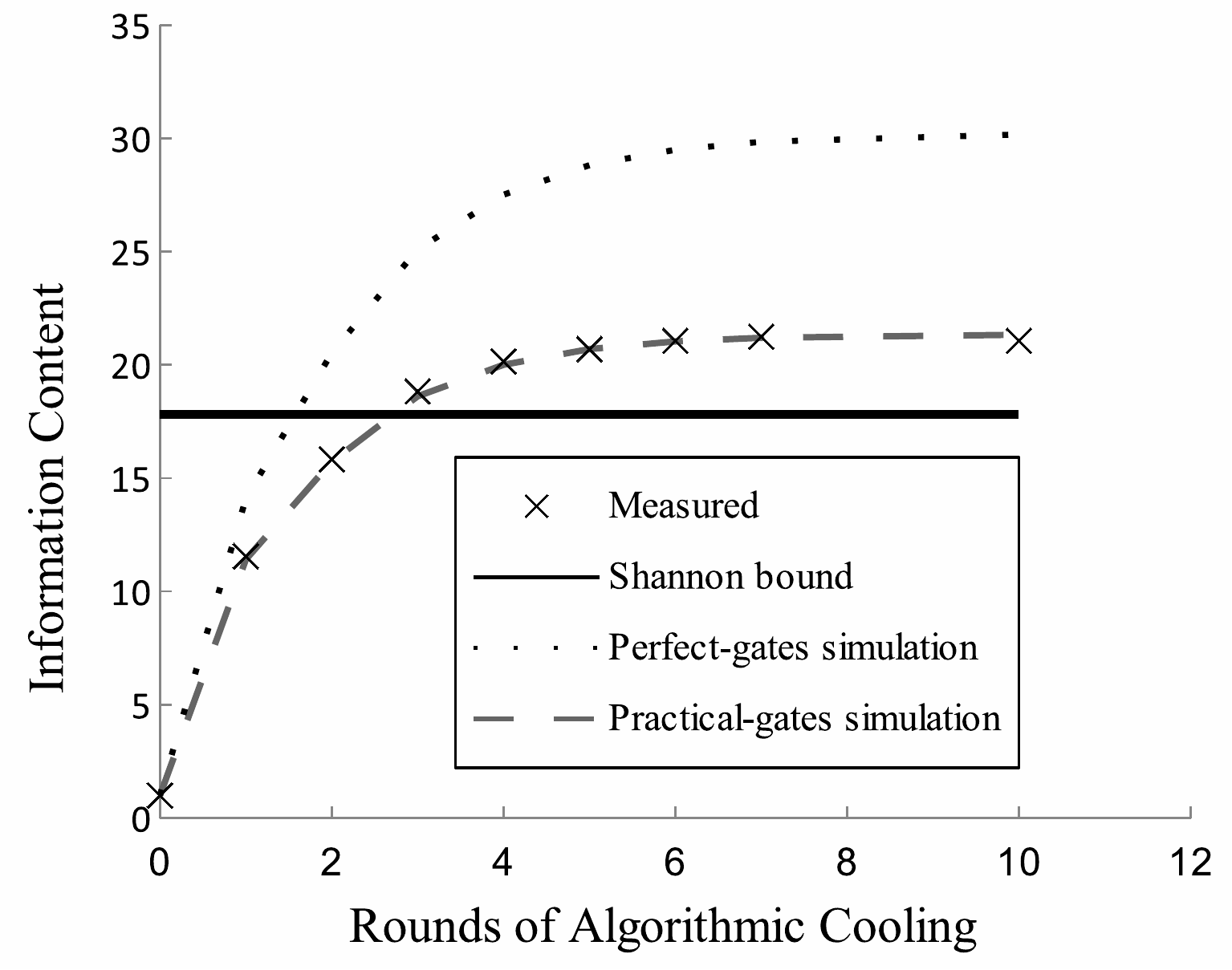}
\end{subfigure}
\begin{subtable} [b] {\columnwidth}
\begin{center}
\begin{tabular} {|c|c|c|}
\hline 
Round & Polarization  & Carbon's $IC$\tabularnewline
\hline
0 & 1.00 & 1.00 $\pm$ 0.04 \tabularnewline
\hline 
1 & 3.40 & 11.56 $\pm$ 0.14\tabularnewline
\hline 
2 & 3.98 & 15.84 $\pm$ 0.16 \tabularnewline
\hline 
3 & 4.34 & 18.84 $\pm$ 0.17 \tabularnewline
\hline 
4 & 4.49 & 20.16 $\pm$ 0.18 \tabularnewline
\hline 
5 & 4.55 & 20.70 $\pm$ 0.18 \tabularnewline
\hline 
6 & 4.59 & 21.07 $\pm$ 0.18\tabularnewline
\hline 
7 & 4.61 & 21.25 $\pm$ 0.18 \tabularnewline
\hline
10 & 4.59 & 21.07 $\pm$ 0.18 \tabularnewline
\hline
\end{tabular}
\end{center}
\vspace{15pt}
\end{subtable}
\caption{On the left, the measured $IC$ vs the simulated $IC$ of C1 at each cooling round of Process 1, in units of $\frac{\eps_{C,eq}^2}{\ln(4)}$ (see~\cite{POTENT}), where $\eps_{C,eq}$ is the carbons' equilibrium polarization. On the right, the measured polarization and $IC$ of C1 in each round. The measured error of all the polarizations is 0.02.}\label{fig:buildup}
\end{figure*}

After seven rounds (see Figure \ref{fig:buildup}), the system reached 
its limit cycle and no more improvement could be expected.
In Process 1, 
C1 was 
cooled by a factor of $4.61\pm0.02$, with $IC_{C1}=21.25\pm0.18$,
significantly higher than 17.84, 
the $IC$ of the entire spin system at equilibrium (see Figures~\ref{fig:buildup} 
and~\ref{fig:Cspectrumbefore+after}). 
Alternatively, we see that the polarization bypassed the information theoretical
bound of $\sqrt{17.84} = 4.22$.
In Process 2 we maximized $IC_{C1,C2}$, 
by adding another delay, D4 (that happened to be equal to D2=5s in the
optimal case), followed by PE. 
We obtained polarizations 
of $3.78\pm0.02$ and $3.4\pm 0.02$ (of C1 and C2 respectively), with 
$IC_{C1,C2}=25.9\pm0.2$. 
In Process 3 we maximized the total $IC$, using an additional delay D5=6s  
before the measurement (in addition, D4 was modified to 6s).
The measured polarizations were 
$2.87\pm0.02$, $2.64 \pm 0.02$ and $3.58 \pm 0.02$ (for C1, C2, and H respectively), with $IC$ of
$28.0\pm0.20$.

\begin{figure}
\centering
Carbons' Information Content Buildup
\includegraphics[width=\columnwidth]{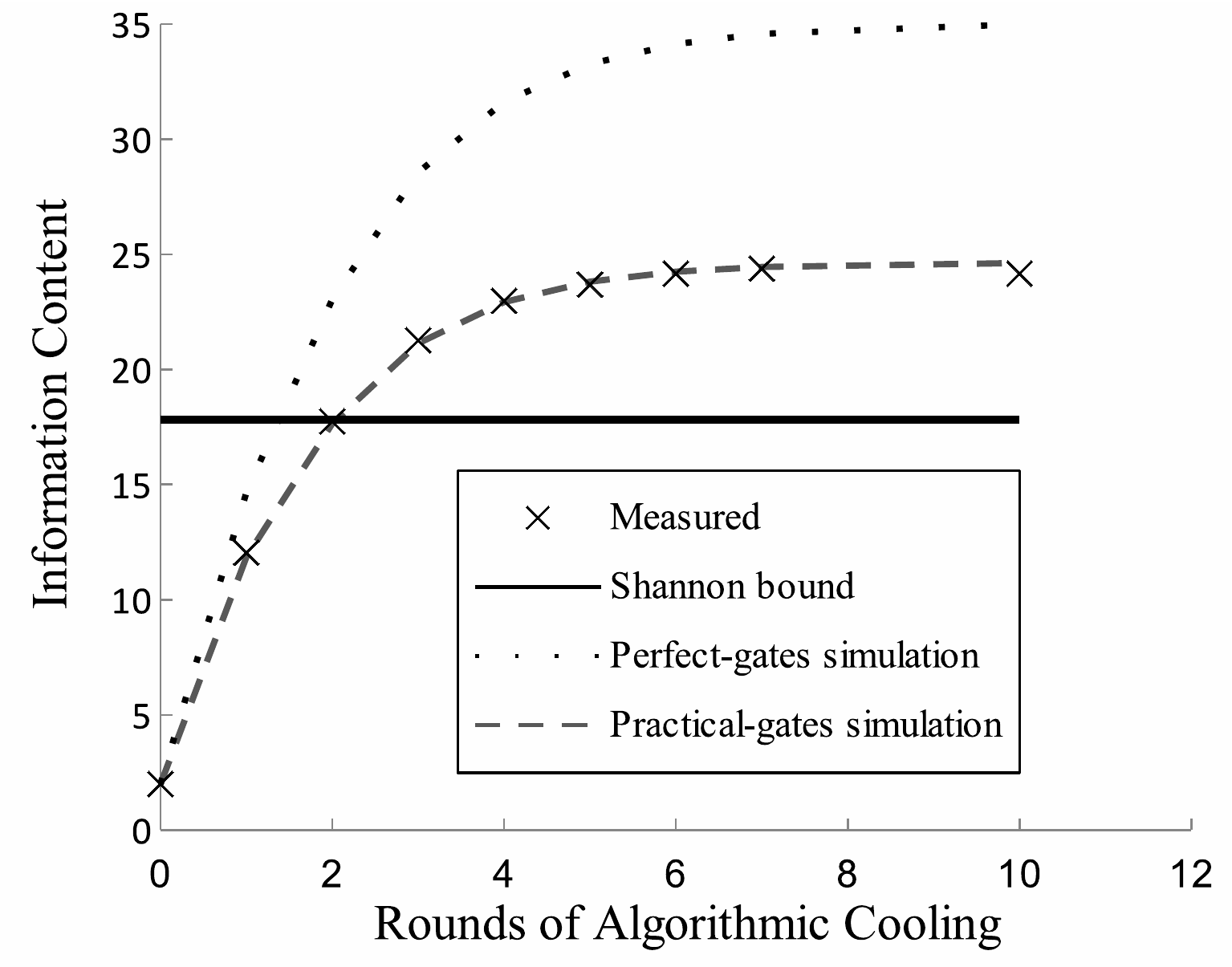}
\caption{The carbons' information content at each cooling round of Process 2 in units of $\frac{\eps_{C,eq}^2}{\ln(4)}$ (see~\cite{POTENT}), where $\eps_{C,eq}$ is
 the equilibrium polarization of the carbons. The measured error of all polarizations is 0.02.}
\label{fig:buildup-of-C1-and-C2}
\end{figure}

There is a gap between SIMPSON's very high predicted pulse efficiency 
and the lab results (see~\cite{AEMW14}). 
The polarization of C2 following a PE pulse was $\approx 3.8$ (0.95 efficiency), 
and a COMP pulse applied on equilibrium state resulted with $\eps_{C1}\approx 2.8$ (0.92 efficiency). 
The main error factors are the hardware's imperfection distorting 
the pulses~\cite{BMW91}, and the pulses themselves being prepared without taking into account the $T_2^*$-relaxation (see table \ref{tbl:TCET1T2}) during the system evolution. The first factor could be negated using a technique mentioned in \cite{RMBL08}. 
Once the measured gate-efficiencies of 0.95 and 0.92 are added into the simulation, the simulated $IC$ per round
fit near-perfectly the measured $IC$, 
see Figures~\ref{fig:buildup} and \ref{fig:buildup-of-C1-and-C2}. 
\begin{table} 
\caption{Measured relaxation times  of TCE in units of seconds.}
{\begin{tabular}{@{}cccc@{}} \toprule
& H & C2 & C1 \\ \colrule
T$_1$ & $2.67\pm 0.03$ & $17.3\pm 0.2$  & $29.2\pm 0.1$ \\
T$_{2}^*$ & $0.2\pm 0.01$ & $0.44\pm 0.03$ & $0.23 \pm0.01$ \\ \botrule
\end{tabular}}
\label{tbl:TCET1T2}
\end{table}

%---------------------------------------

\section{Summary} 
Using optimal control, we demonstrated the first single-round and multiple round AC 
applied on liquid state NMR. We bypassed Shannon's bound in three different
processes. 
The current optimal control methods (GRAPE), and better ones such as a second order GRAPE~\cite{FSGK11} 
and Krotov based optimization~\cite{MTZN+08} could enable various applications
of AC in magnetic resonance spectroscopy~\cite{AAC-pat, EGMW11,BEMW14} and maybe
also other
potential applications~\cite{RMM07,HRM07,WHRSM08,BP08,LPS10,CML11, Renner12,Blank13,XYX+14,Lloyd14}.

\section{Acknowledgments}
We thank Prof. Asher Schmidt and Dr. Yael Balasz for enlightening discussions. This work was supported in part by the Wolfson Foundation and the Israeli MOD Research and Technology Unit. YE thanks the Institut Transdisciplinaire d'Information Quantique.\@

\bibliography{mine}

\end{document}